\documentclass[%
  reprint,
]{revtex4-2}

\pdfoutput=1

\usepackage{amsmath}
\usepackage{amssymb}
\usepackage{bm,bbm}
\usepackage{mathrsfs}

\usepackage{graphicx}

\usepackage{enumitem}

\usepackage[bookmarks=false]{hyperref}
\hypersetup{colorlinks=true,citecolor=blue,linkcolor=red,%
urlcolor=blue,pdfstartview=FitH,bookmarksopen=true}

\usepackage[T1]{fontenc}
\usepackage[osf,sc]{mathpazo}

\usepackage{amsthm}
\newtheoremstyle{note}
{3pt}
{3pt}
{}
{}
{\itshape}
{:}
{.5em}
{}

\usepackage{physics}

\newtheorem{theorem}{Theorem}

\newtheorem{corollary}[theorem]{Corollary}

\DeclareMathOperator{\diag}{diag}

\newcommand{\bs}{\bm{s}}
\newcommand{\ee}{\mathrm{e}}
\newcommand{\ii}{\mathrm{i}}

\newcommand{\cE}{\mathcal{E}}
\newcommand{\cF}{\mathcal{F}}

\newcommand{\cH}{\mathcal{H}}

\newcommand{\cP}{\mathcal{P}}

\newcommand{\cT}{\mathcal{T}}

\usepackage{dsfont}

\newcommand{\I}{\mathds{1}}

\newcommand{\hG}{\hat{\Gamma}}
\newcommand{\htG}{\hat{C}}
\newcommand{\hU}{\hat{U}}
\newcommand{\hD}{\hat{D}}
\newcommand{\hF}{\hat{F}}
\newcommand{\sP}{P}

\newcommand{\bmid}{:}

\begin{document}

\title{Evolution Operator Can Always be Separated into the Product of\\
Holonomy and Dynamic Operators}

\author{Xiao-Dong Yu}
\email{yuxiaodong@sdu.edu.cn}
\affiliation{Department of Physics, Shandong University, Jinan 250100, China}
\author{D. M. Tong}
\email{tdm@sdu.edu.cn}
\affiliation{Department of Physics, Shandong University, Jinan 250100, China}

\date{\today}

\begin{abstract}

The geometric phase is a fundamental quantity characterizing the holonomic
feature of quantum systems. It is well known that the evolution operator of
a quantum system undergoing a cyclic evolution can be simply written as the
product of holonomic and dynamical components for the three special cases
concerning the Berry phase, adiabatic non-Abelian geometric phase, and
nonadiabatic Abelian geometric phase. However, for the  most general case
concerning the nonadiabatic non-Abelian geometric phase, how to separate the
evolution operator into holonomic and dynamical components is a long-standing
open problem.
In this work, we solve this open problem. We show that the evolution operator
of a quantum system can always be separated into the product of holonomy and
dynamic operators. Based on it, we further derive a matrix representation of
this separation formula for cyclic  evolution, and give a necessary and
sufficient condition for a general evolution being purely holonomic. Our
finding is not only of theoretical interest itself, but also of vital
importance for the application of quantum holonomy. It unifies the
representations of all four types of evolution  concerning the
adiabatic/nonadiabatic Abelian/non-Abelian geometric phase, and provides
a general approach to realizing purely holonomic evolution.
\end{abstract}

\maketitle

\textit{Introduction.---}%
%
Nature is replete with phenomena where a quantity fails to return to its
original value although the driving parameters undergo a cyclic evolution.
Holonomy is used to characterize the geometrical essence of such phenomenon, in
which the value difference remains even if the local rate of change is always
zero \cite{Berry1990}. This kind of holonomic effect plays crucial roles in
various fields of physics \cite{Shapere.Wilczek1989}, including, for example,
in mechanics the change of the swing plane of a Foucault pendulum after one
rotation of the earth, in optics the change in the direction of linear
polarization of light along a coiled optical fiber, and in general relativity
the change of reference frames around a closed loop in spacetime.

In the quantum regime, despite of earlier studies in specific systems
\cite{Pancharatnam1956,LonguetHiggins.etal1958,Aharonov.Bohm1959,Stone1976,
Mead.Truhlar1979}, the Berry phase found in 1984 \cite{Berry1984} is often
considered as the seminal theory on the quantum holonomic phenomenon. This
finding represents a special case of quantum holonomy, the adiabatic Abelian
geometric phase. The generalization of the Berry phase to adiabatic
non-Abelian geometric phase \cite{Wilczek.Zee1984},
nonadiabatic Abelian geometric phase \cite{Aharonov.Anandan1987}, and
nonadiabatic non-Abelian geometric phase \cite{Anandan1988} were soon
established.
The holonomic nature of geometric phases is of broad importance
in various research fields, such as in condensed matter physics
\cite{Xiao.etal2010}, in quantum chemistry \cite{Bohm.etal2003}, in quantum
field theory \cite{Witten1989,Nayak.etal2008}, in quantum gravity
\cite{Rovelli2008,Ashtekar.Bianchi2021}, and in quantum information
\cite{Zhang.etal2023}. Moreover, this kind holonomic nature has also been
experimentally observed and manipulated in a variety of physical platforms,
such as in superconductors \cite{Leek.etal2007,AbdumalikovJr.etal2013}, in
nitrogen-vacancy centers \cite{Zu.etal2014,Yale.etal2016,Zhou.etal2017}, in
trapped ions \cite{Leibfried.etal2003}, in molecular ensembles
\cite{Jones.etal2000,Feng.etal2013}, and in photonic systems
\cite{Yang.etal2019,Neef.etal2023}.

For a quantum system undergoing a cyclic evolution, the state difference, i.e.,
the evolution operator transforming the initial state to the final state, is
not purely holonomic in general, but it can be simply written as the product of
the holonomic and dynamical components for the three special cases concerning
the Berry phase, adiabatic non-Abelian geometric phase, and nonadiabatic
Abelian geometric phase. However, the situation is different for the most
general case concerning the nonadiabatic non-Abelian geometric phase. In all
the previous works on this issue, the holonomic component is blended with the
dynamical component, and the evolution operator cannot be separated into the
product of them except for some special cases such that they commute with each
other.

How to separate the evolution operator of a general quantum system into the
product of holonomic and dynamical components has been a long-standing open
problem ever since the discovery of nonadiabatic quantum holonomy. The
difficulty comes from the noncommutativity of the holonomic and dynamical
components, both of which are related to the time-ordered integral.
Solving this problem is not only of theoretical interest itself, but also of
vital importance for the application of quantum holonomy.  For example,
holonomy-based quantum computation and quantum control relay on the separation
of holonomic and dynamical components, which ensures the possibility of
eliminating the dynamical component from cyclic evolution and hence taking full
advantage of the holonomy against control errors
\cite{Ekert.etal2000,Wang.Matsumoto2001,Zhu.Wang2002,
Zanardi.Rasetti1999,Sjoeqvist.etal2012,Xu.etal2012}.

In this work, we solve this open problem. We first show that the evolution
operator of a quantum system can always be separated into the product of
holonomy and dynamic operators. Based on it, we further derive a matrix
representation of this separation formula for cyclic evolution, and give
a necessary and sufficient condition for a general evolution being purely
holonomic. Our finding unifies the representations of all four types of
evolution concerning the adiabatic/nonadiabatic Abelian/non-Abelian geometric
phase, and provides a general approach to realizing purely holonomic
evolution.

\textit{Preliminaries.---}%
%
We start by recalling the notion of quantum holonomy. Since Abelian geometric
phase can be taken as a special case of non-Abelian geometric phase, our
discussion focuses on non-Abelian geometric phase, to which we refer as quantum
holonomy or holonomy for simplicity.

Quantum holonomy was first studied in adiabatic evolution as a generalization
of the Berry phase to degenerate eigenstates \cite{Wilczek.Zee1984}. It arises
in the quantum system governed by a slowly changing Hamiltonian $H(\bs)$ with
a degenerate eigenvalue $E(\bs)$ of order $\ell$, where $\bs=\bs(t)$ is a set
of time-dependent parameters with $\bs(T)=\bs(0)$.
Without ambiguity, we will use $P(\bs)$ to denote both the subspace
spanned by the $\ell$ degenerate eigenstates and the corresponding rank-$\ell$
orthogonal projector. Then,
$P(\bs)=\ketbra{\phi_i(\bs)}:=\sum_{i=1}^\ell\ketbra{\phi_i(\bs)}$, where
$\{\ket{\phi_i(\bs)}\}_{i=1}^\ell$ is an arbitrary orthonormal basis of the
$\ell$-degenerate eigenspace. Here and in the following, the Einstein summation
convention is employed , i.e., repeated indices are implicitly summed over from
$1$ to $\ell$. For a quantum system evolving adiabatically, any state
initially in the subspace $\sP(\bs(0))$ will be in the subspace $\sP(\bs(t))$
at time $t$, and go back to the initial subspace at $t=T$ as
$\sP(\bs(T))=\sP(\bs(0))$.

Let $\ket{\psi_j(t)}$ be the state of the quantum system at time $t$, which is
initially in  $\ket{\psi_j(0)}=\ket{\phi_j(0)}$. Then, $\ket{\psi_j(t)}$ admits
the expression $\ket{\psi_j(t)}=U_{ij}(t)\ket{\phi_i(\bs(t))}$, where
$U(t)=\qty[U_{ij}(t)]_{i,j=1}^\ell$ is the transformation matrix between the
basis and the states, and $U(T)$ gives the evolution operator after a cyclic
evolution. By substituting $\ket{\psi_j(t)}$ into the Schr\"odinger equation,
one can obtain that $U(t)=\ee^{-\ii\int_0^tE(\bs(\tau))\dd\tau}\Gamma(\bs(t))$
with $\Gamma(\bs)=\cP\exp\qty[\int\bm{A}(\bm\varsigma)\cdot\dd{\bm\varsigma}]$,
where $\cP$ denotes the path ordering along the curve $\qty{\bs(\tau)\bmid 0\le
\tau \le t}$, and $\bm{A}(\bs)=[\bm{A}_{ij}(\bs)]_{i,j=1}^\ell$ with
$\bm{A}_{ij}(\bs)=-\bra{\phi_i(\bs)}\grad_{\bs}\ket{\phi_j(\bs)}$.  Thus, after
a cyclic evolution with period $T$, the evolution operator acting on the
initial subspace reads $U(T)=\ee^{-\ii\int_0^TE(\bs(\tau))\dd\tau}\Gamma(T)$,
where $\Gamma(T)=\cP\exp\qty[\oint \bm{A}(\bs)\cdot\dd{\bs}]$ is the quantum
holonomy in adiabatic evolution. $\Gamma(T)$ is gauge invariant in the
sense that it depends only on the subspace $P(\bs)$ but not on the choice of
the basis $\{\ket{\phi_i(\bs)}\}_{i=1}^\ell$,
as long as $\ket{\phi_i(T)}=\ket{\phi_i(0)}=\ket{\psi_i(0)}$.

Quantum holonomy can also be generalized to the nonadiabatic case
\cite{Anandan1988}.
Consider a $d$-dimensional quantum system governed by Hamiltonian $H(t)$.
If there exists a set of the orthonormal states
$\{\ket{\psi_i(t)}\}_{i=1}^\ell$  satisfying the Schr\"odinger equation
$\ii\ket{\dot\psi_i(t)}=H\ket{\psi_i(t)}$ and the cyclic evolution condition
$\ketbra{\psi_i(T)}=\ketbra{\psi_i(0)}$, i.e.,
$\sum_{i=1}^\ell\ketbra{\psi_i(T)}=\sum_{i=1}^\ell\ketbra{\psi_i(0)}$,
with $T$ being the evolution period, then one can define the quantum holonomy
for the subspace $\sP(t)$ spanned by $\{\ket{\psi_i(t)}\}_{i=1}^\ell$
similarly to the adiabatic case,
\begin{equation}
  \Gamma(T)=\cP\exp\qty[\int_{0}^TA(t)\dd{t}],
  \label{eq:holonomy}
\end{equation}
where $A(t)=[A_{ij}(t)]_{i,j=1}^\ell$ with
\begin{equation}
  A_{ij}(t)=\braket{\dot\phi_i(t)}{\phi_j(t)}
\label{eq:defA}
\end{equation}
is an anti-Hermitian matrix. Here, $\{\ket{\phi_i(t)}\}_{i=1}^\ell$ satisfying
$\ket{\phi_i(T)}=\ket{\phi_i(0)}=\ket{\psi_i(0)}$
is an arbitrary basis of the subspace $\sP(t)$,
and the dot ($\,\dot{}\,$) denotes the derivative with respect to $t$.  Again,
$\Gamma(T)$ does not depend on any special choice of the basis, being gauge
invariant.

A critical problem existing in all the previous works on the nonadiabatic
quantum holonomy is that the evolution operator acting on the subspace cannot
be separated into the product of holonomic and dynamical components. To see
this clearly, we recall the processing procedure in the literature (see, e.g.,
\cite{Anandan1988}). The state $\ket{\psi_j(t)}$ is expressed as
\begin{equation}
  \ket{\psi_j(t)}=U_{ij}(t)\ket{\phi_i(t)}.
 \label{eq:defU}
\end{equation}
Still, $U(t)=\qty[U_{ij}(t)]_{i,j=1}^\ell$ represents the transformation matrix
between the basis and the states, and $U(T)$ gives the evolution operator after
a cyclic evolution. Substituting  $\ket{\psi_j(t)}$ into the Schr\"odinger
equation gives the differential equation,
\begin{equation}
  \dv{t}U(t)=\qty[A(t)+K(t)]U(t),
  \label{eq:inseparable}
\end{equation}
where $K(t)=[K_{ij}(t)]_{i,j=1}^\ell$ with
$K_{ij}(t)=-\ii\mel{\phi_i(t)}{H(t)}{\phi_j(t)}$.
Equation~\eqref{eq:inseparable} implies that
$U(T)=\cT\exp\qty[\int_0^T\qty[A(t)+K(t)]\dd{t}]$ with $\cT$ denoting the time
ordering. Clearly, although $U(T)$ can be written as the product of
$\cT\exp\qty[\int_0^T A(t)\dd{t}]$ and $\cT\exp\qty[\int_0^T K(t)\dd{t}]$ in
the special case that $[A(t_1),K(t_2)]=0$ for any $t_1$ and $t_2$, this
separation is invalid in general. Therefore, one cannot simply separate the
holonomic component  $\Gamma(T)$ from the dynamical component by starting from
the known
differential equation~\eqref{eq:inseparable}. To realize the
separation, we need to construct a new differential equation.

\textit{Main results.---}%
%
With the above preliminaries, we can now present our results. We first
construct a new differential equation of the evolution operator, based on
which we derive a universal formula for separating the evolution operator into
the product of holonomy and dynamic operators.

The evolution operator acting on the subspace spanned by
$\qty{\ket{\psi_i(0)}}_{i=1}^\ell$ plays the role that it transforms an
arbitrary state in the initial subspace $\sP(0)$ to a corresponding state in
the subspace $\sP(t)$ at time $t$. For example, if the quantum system is
initially in the superposition $c_i\ket{\psi_i(0)}$, it will be in the state
$c_i\ket{\psi_i(t)}$ at time $t$.  Thus, the evolution operator acting on the
subspace can be written as
\begin{equation}
  \hU(t)=\ketbra{\psi_j(t)}{\psi_j(0)}.
  \label{eq:defUhat}
\end{equation}
Here, we use the hat ($\,\hat{}\,$) to emphasize that $\hU(t)$ is an operator,
which should not be simply viewed as an $\ell\times\ell$ matrix.
The evolution operator $\hU(t)$ is different from the transformation matrix
$U(t)$ defined in Eq.~\eqref{eq:defU} in general, but coincides with $U(t)$
at time $T$.

Based on the expression \eqref{eq:defUhat}, we can construct the following
differential equation satisfied by the evolution operator
\footnote{
This can be easily verified by expanding the two terms in the
right-hand side of Eq.~\eqref{eq:separationDifferential} as
$\dot{P}(t)\hU(t)=\ketbra*{\dot\psi_j(t)}{\psi_j(0)}
-F_{ij}(t)\ketbra*{\psi_i(t)}{\psi_j(0)}$ and
$\hU(t)\hF(t)=F_{ij}(t)\ketbra*{\psi_i(t)}{\psi_j(0)}$.
}
\begin{equation}
  \dv{t}\hU(t)=\dot{P}(t)\hU(t)+\hU(t)\hF(t),
  \label{eq:separationDifferential}
\end{equation}
where
\begin{equation}
  \hF(t)=F_{ij}(t)\ketbra{\psi_i(0)}{\psi_j(0)},
  \label{eq:defFOperator}
\end{equation}
with
\begin{equation}
  F_{ij}(t)=-\braket{\dot\psi_i(t)}{\psi_j(t)}
  =-\ii\mel{\psi_i(t)}{H(t)}{\psi_j(t)}.
  \label{eq:defF}
\end{equation}

It is interesting to note that the form of
Eq.~\eqref{eq:separationDifferential} is different from
Eq.~\eqref{eq:inseparable} in the sense that $\dot{P}(t)$ and $\hF(t)$ appear
on different sides of $\hU(t)$, while their counterparts $A(t)$ and $K(t)$
appear on the same side of $U(t)$. It is exactly this subtle difference that
makes the separation possible. Mathematically, if $\dot{X}(t)=L(t)X(t)$ and
$\dot{Y}(t)=Y(t)R(t)$, then $Z(t)=X(t)Y(t)$ satisfies that
$\dot{Z}(t)=L(t)Z(t)+Z(t)R(t)$.  Note also that $P(0)$ is the identity operator
on the initial subspace, and so is $\hU(0)$. Thus,
Eq.~\eqref{eq:separationDifferential} implies the following separation formula
in operator form.

\begin{theorem}
  Let $P(t)$ be a subspace spanned by $\ell$ orthonormal states
  $\qty{\ket{\psi_i(t)}}_{i=1}^\ell$ of a quantum system with Hamiltonian
  $H(t)$, then the evolution operator $\hU(t)$ acting on the subspace can
  always be separated into the product of the holonomy operator $\hG(t)$ and
  dynamic operator $\hD(t)$,
  \begin{equation}
    \hU(t)=\hG(t)\hD(t)
    \label{eq:separation}
  \end{equation}
  with
  \begin{align}
    \label{eq:holonomyOperatorIntegral}
    \hG(t)&=\cP\exp\qty[\int_{0}^t\dot{P}(\tau)\dd{\tau}]P(0),\\
    \label{eq:defDOperator}
    \hD(t)&=P(0)\bar\cT\exp\qty[\int_{0}^t\hF(\tau)\dd{\tau}],
  \end{align}
  where $P(t)=\ketbra{\psi_i(t)}$,
  $\hF(t)= F_{ij}(t)\ketbra{\psi_i(0)}{\psi_j(0)}$,
  $F_{ij}(t)=-\ii\mel{\psi_i(t)}{H(t)}{\psi_j(t)}$, and $\cP$ and $\bar\cT$ are
  the path ordering and reverse time ordering,
  respectively.
  \label{thm:separationOperator}
\end{theorem}

Alternatively, one can express
Eqs.~(\ref{eq:holonomyOperatorIntegral},\,\ref{eq:defDOperator})
in differential form
\begin{align}
  \dv{t}\hG(t)=\dot{P}(t)\hG(t),\quad
  \dv{t}\hD(t)=\hD(t)\hF(t),
\end{align}
with the initial conditions that $\hG(0)=\hD(0)=P(0)$. In
Eq.~\eqref{eq:holonomyOperatorIntegral}, the time ordering $\cT$ has been
replaced by the path ordering $\cP$, as $\hG(t)$ is independent of the
evolution details such as the evolution rate but only depends on the path of
the $\ell$-dimensional subspaces $\qty{\sP(\tau)\bmid 0\le\tau\le t}$. Hence,
$\hG(t)$ is a holonomic component, and thus we call it the holonomy operator
for the evolution.
The other component $\hD(t)$ depends on the dynamical details of the evolution,
hence we call it the dynamic operator for the evolution. Besides, $\hF(t)$
defined by Eq.~\eqref{eq:defFOperator} is an anti-Hermitian
operator, i.e., $\hF^\dagger(t)=-\hF(t)$, and therefore the Hermitian conjugate
of the dynamic operator can be written in the time ordering form as
\begin{equation}
  \hD^\dagger(t)=\cT\exp\qty[-\int_{0}^t\hF(\tau)\dd{\tau}]P(0).
  \label{eq:defDdaggerOperator}
\end{equation}

So far, we have proved that the evolution operator can always be
separated into the product of holonomy and dynamic operators as in
Theorem~\ref{thm:separationOperator}. In the following, we will apply the
theorem to the cyclic evolution, and give the matrix representation of the
separation formula.

To this end, we need to express operators $\hU(T)$, $\hD(T)$, and $\hG(T)$
with their corresponding matrices.  By using Eq.~\eqref{eq:defU},
the evolution operator $\hU(t)$ can be written as
$\hU(t)=U_{ij}(t)\ketbra{\phi_i(t)}{\phi_j(0)}$, which coincides with the
transformation matrix $U(t)$ at time $T$, i.e.,
\begin{equation}
  \hU(T)=U_{ij}(T)\ketbra{\phi_i(0)}{\phi_j(0)}.
  \label{eq:UT}
\end{equation}
Here, $\{\ket{\phi_i(t)}\}_{i=1}^\ell$ satisfying
$\ket{\phi_i(T)}=\ket{\phi_i(0)}=\ket{\psi_i(0)}$ is still used to denote
a basis of the subspace $\sP(t)$.

In analogy to Eq.~\eqref{eq:UT}, we can derive the matrix
representation of the dynamic operator $\hD(T)$ from its definition.
Equations~(\ref{eq:defFOperator},\,\ref{eq:defDOperator}) directly imply that
\begin{equation}
  \hD(t)=D_{ij}(t)\ketbra{\phi_i(0)}{\phi_j(0)}
  \label{eq:DMatrixForm}
\end{equation}
with $D(t)=\bar{\cT}\exp\qty[\int_{0}^tF(\tau)\dd{\tau}]$, where
$F(t)=\qty[F_{ij}(t)]_{i,j=1}^\ell$ is an anti-Hermitian matrix defined by
Eq.~\eqref{eq:defF}. Equation (\ref{eq:DMatrixForm}) gives
$\hD(T)=D_{ij}(T)\ketbra{\phi_i(0)}{\phi_j(0)}$ at the time $T$.

The remaining task is to show that $\hG(T)$ coincides with the holonomy matrix
$\Gamma(T)$ defined in Eq.~\eqref{eq:holonomy} for cyclic evolution. For this,
we first use $\htG(t)$ to denote $\Gamma_{ij}(t)\ketbra{\phi_i(t)}{\phi_j(0)}$,
where $\Gamma(t)=\qty[\Gamma_{ij}(t)]_{i,j=1}^\ell
=\cT\exp\qty[\int_{0}^tA(\tau)\dd{\tau}]$ with
$A(t)=\qty[\braket{\dot\phi_i(t)}{\phi_j(t)}]_{i,j=1}^\ell$, and will then
prove $\hG(t)=\htG(t)$ by demonstrating that they satisfy the same differential
equation with the same initial condition.  From the definitions of $\htG(t)$
and $\Gamma(t)$, we have
\begin{equation}
  \dv{t}\htG(t)
  =\dot\Gamma_{ij}(t)\ketbra{\phi_i(t)}{\phi_j(0)}+
  \Gamma_{ij}(t)\ketbra{\dot\phi_i(t)}{\phi_j(0)}
  \label{eq:holonomyOperatorDT}
\end{equation}
with $\dot\Gamma_{ij}(t)=A_{ik}(t)\Gamma_{kj}(t)
=\braket{\dot\phi_i(t)}{\phi_k(t)}\Gamma_{kj}(t)$.
Simple calculations show that the first term on the right-hand side of
Eq.~\eqref{eq:holonomyOperatorDT} reduces to
$\ketbra{\phi_i(t)}{\dot\phi_i(t)}\htG(t)$ and the second term reduces to
$\ketbra{\dot\phi_i(t)}{\phi_i(t)}\htG(t)$.  Therefore, $\htG(t)$
satisfies the differential equation $\dv{t}\htG(t)=\dot{P}(t)\htG(t)$ with
the initial condition that $\htG(0)=P(0)$, which is the same to $\hG(t)$. Thus,
we prove that $\hG(t)=\htG(t)$, i.e.,
\begin{equation}
  \hG(t)=\Gamma_{ij}(t)\ketbra{\phi_i(t)}{\phi_j(0)},
  \label{eq:holonomyOperatorT}
\end{equation}
which gives $\hG(T)=\Gamma_{ij}(T)\ketbra{\phi_i(0)}{\phi_j(0)}$ at the time $T$.

With Eqs.~(\ref{eq:separation},\,\ref{eq:UT},\,\ref{eq:DMatrixForm},
\ref{eq:holonomyOperatorT}), we obtain the following
separation formula for cyclic evolution in matrix form.

\begin{theorem}
  If the subspace $P(t)$ spanned by  $\ell$ orthonormal states
  $\qty{\ket{\psi_i(t)}}_{i=1}^\ell$ evolves cyclically with period $T$, i.e.,
  $P(T)=P(0)$, then the evolution operator acting on the subspace at time $T$
  has the matrix representation,
  \begin{equation}
    U(T)=\Gamma(T)D(T),
    \label{eq:separationMatrix}
  \end{equation}
  where $\Gamma(T)=\cP\exp\qty[\int_0^TA(t)\dd{t}]$ and
  $D(T)=\bar{\cT}\exp\qty[\int_{0}^TF(t)\dd{t}]$ are the holonomic and
  dynamical components, respectively.
  \label{thm:separation}
\end{theorem}

Before proceeding further, we would like to
add a few remarks.  First, our finding unifies the representations of all four
types of evolution concerning the adiabatic/nonadiabatic Abelian/non-Abelian
geometric phase.
In the adiabatic Abelian/non-Abelian case, $F_{ij}(t)=-\ii E(t)\delta_{ij}$,
and thus Eq.~\eqref{eq:separationMatrix} reduces to
$U(T)=\ee^{-\ii\int_0^TE(t)\dd{t}}\Gamma(T)$, which is just the
well-known results of Berry \cite{Berry1984} (when $\ell=1$)
and Wilczek and Zee \cite{Wilczek.Zee1984} (when $\ell\ne 1$).
In the nonadiabatic Abelian case, i.e., $P(t)=\ketbra{\psi(t)}$,
Eq.~\eqref{eq:separationMatrix} reduces to the celebrated formula of
Aharonov and Anandan \cite{Aharonov.Anandan1987},
$\gamma(T)=\arg\braket{\psi(0)}{\psi(T)}
+\int_{t=0}^T\mel{\psi(t)}{H(t)}{\psi(t)}\dd{t}$.
Second, Eqs.~\eqref{eq:separation} and \eqref{eq:separationMatrix}
provide separation formulae in the operator and matrix forms, respectively, but
Eq.~\eqref{eq:separation} is more general than Eq.~\eqref{eq:separationMatrix}.
The operator form \eqref{eq:separation} holds for any time $t$, or
equivalently, it also holds for noncyclic evolution, while the matrix form
\eqref{eq:separationMatrix} holds only for cyclic evolution.
Third, a fundamental difference between $\hG(t)$ and $\Gamma(t)$
is that the gauge invariance holds
at any time $t$ for the former but only at time $T$ for the latter.
Moreover, the gauge-invariant quantity $\hG(t)$ gives the operator form of the
so-called parallel transport, i.e., it satisfies that
$\hG^\dagger(t)\dv{t}\hG(t)=0$
\footnote{This follows from that $\dv{t}\hG(t)=\dot{P}(t)\hG(t)$,
$P(t)\hG(t)=\hG(t)$, and $P(t)\dot{P}(t)P(t)=0$.}.


\textit{Purely holonomic evolution.---}%
%
A crucial issue for the application of quantum holonomy
is to determine when a quantum evolution is purely holonomic. Explicitly, we
call a cyclic evolution purely holonomic if $U(T)$ is equal to $\Gamma(T)$ up
to a global phase, i.e., $U(T)=\ee^{\ii\alpha}\Gamma(T)$ for some real number
$\alpha$. We note that this is different from Abelian geometric phases, in the
applications of which two or more paths are considered and thus the phases, or
rather, the difference of the phases matters. From
Theorem~\ref{thm:separation}, one can directly obtain the following necessary
and sufficient condition for purely holonomic evolution.

\begin{corollary}
  If  the subspace $P(t)$ spanned by  $\ell$ orthonormal states
  $\qty{\ket{\psi_i(t)}}_{i=1}^\ell$ evolves cyclically with period $T$, i.e.,
  $P(T)=P(0)$, then
  the evolution is purely holonomic if and only if
  \begin{equation}
    D^\dagger(T)=\cT\exp\qty[
    -\int_{0}^TF(t)\dd{t}]=
    \ee^{\ii\alpha}\I_\ell,
    \label{eq:purelyHolonomic}
  \end{equation}
  where $\alpha$ is some real number, $\I_\ell$ is the $\ell\times\ell$
  identity matrix, and $F(t)=\qty[F_{ij}(t)]_{i,j=1}^\ell$ with
  $F_{ij}(t)=-\ii\mel{\psi_i(t)}{H(t)}{\psi_j(t)}$.
  \label{thm:purelyHolonomic}
\end{corollary}

Corollary~\ref{thm:purelyHolonomic} shows that for a quantum system, the
evolution operator acting on the subspace $\sP(0)$ is purely holonomic if
and only if the Hamiltonian governing the quantum satisfies
Eq.~(\ref{eq:purelyHolonomic}). This provides a general approach for realizing
purely holonomic evolution. Specially, this opens a new avenue for holonomic
quantum computation and holonomic quantum control. Note that in the previous
schemes, either the systems must be in
adiabatic evolution \cite{Zanardi.Rasetti1999} or satisfy the parallel
transport condition \cite{Sjoeqvist.etal2012,Xu.etal2012}, which are just
corresponding to the special cases of Corollary~\ref{thm:purelyHolonomic} with
$F(t)=-iE(t)\I_\ell$ and  $F(t)=0$, respectively.  Yet, our result shows that
these requirements are unnecessary for purely holonomic evolution.
This largely extends the applicability of holonomic quantum computation.
In the Supplemental Material, 
we take a one-parameter Hamiltonian, which is widely used in holonomic
quantum computation, as a concrete example to illustrate this point.

Besides, the use of Corollary~\ref{thm:purelyHolonomic} can be flexible.
For a cyclic evolution
$\qty{\ket{\psi_i(t)}}_{i=1}^\ell$ ($0\le t\le T_1$) with the dynamical
component $D(T_1)\not\propto\I_\ell$, it is possible to construct an adjacent
cyclic evolution $\qty{\ket{\psi_i(t)}}_{i=1}^\ell$ ($T_1\le t\le T_1+T_2$)
such that $D^\dagger(T_1+T_2;T_1)\propto D(T_1)$, where $D^\dagger(T_1+T_2;T_1)
=\cT\exp\qty[-\int_{T_1}^{T_1+T_2}F(t)\dd{t}]$. In this way, the dynamical
components cancel out and only the holonomic components remain.
This strategy works because we can regard $\qty{\ket{\psi_i(t)}}_{i=1}^\ell$
($0\le t\le T_1+T_2$) as an overall cyclic evolution, whose dynamical component
satisfies $D^\dagger(T_1+T_2)=D^\dagger(T_1+T_2;T_1)D^\dagger(T_1)$.

\textit{Conclusions.---}%
%
We have shown that the evolution operator of a quantum system can always
be separated into the product of holonomy and dynamic operators, of which
the operator expression and the matrix representation are stated as
Theorem~\ref{thm:separationOperator} and Theorem~\ref{thm:separation},
respectively.  From the fundamental perspective, our finding solves
a long-standing open problem in the study of quantum holonomy, and unifies the
representations of all four types of evolution concerning the
adiabatic/nonadiabatic Abelian/non-Abelian geometric phase. From the practical
perspective, our separation formula provides a general approach for realizing
purely holonomic evolution, which can find widespread applications in quantum
information and quantum control.  For example, our approach
can largely extend the applicability of holonomic quantum computation.

For the future research, it would be very interesting to apply our approach for
implementing holonomic quantum computation and holonomic quantum control in
actual quantum systems, both theoretically and experimentally.
Furthermore, as
quantum holonomy is a fundamental geometric quantity in quantum physics, our
result may also lead to deeper understandings of the geometric phenomena in
various fields.  For example, as our approach holds for any quantum evolution,
it opens a more flexible avenue for simulating the non-Abelian gauge field.
This may shed light on the investigation of the factional quantum Hall
effect \cite{Semenoff.Sodano1986}, lattice gauge theory
\cite{Banuls.Cichy2020}, topological field theory
\cite{Witten1989,Nayak.etal2008}, and loop quantum gravity
\cite{Rovelli2008,Ashtekar.Bianchi2021}, in various quantum simulation
platforms \cite{Georgescu.etal2014,Daley.etal2022}.

  This work was supported by
  the National Natural Science Foundation of China
  (Grants No. 12174224 and No.  12205170)
  and the Shandong Provincial Natural Science Foundation of China
  (Grant No. ZR2022QA084).

\twocolumngrid

%


\onecolumngrid
\bigskip\bigskip\bigskip
\appendix*
\renewcommand{\theequation}{S\arabic{equation}}
\setcounter{equation}{0}

{\centering \large \bfseries Supplemental Material\\[2em]}

Let us consider the one-parameter Hamiltonian,
\begin{equation}
  H(t)=\omega(t)\cH,
\end{equation}
where $\cH$ is time-independent. For simplicity, we restrict ourselves to
three-dimensional systems, which is actually the most widely used form of
Hamiltonians for implementing single-qubit gates in nonadiabatic holonomic
quantum computation \cite{Zhang.etal2023}. The generalization to
high-dimensional systems is straightforward.

Now, our task is to find a two-dimensional cyclic evolution
$\qty{\ket{\psi_1(t)},\ket{\psi_2(t)}}$ ($0\le t\le T$), which is purely
holonomic. According to Corollary~\ref{thm:purelyHolonomic}, this is equivalent
to the following two conditions:
\begin{align}
  \label{eqa:cond1}
  \ket{\psi_0(T)}=&\ee^{-\ii
  \cH\theta_T}\ket{\psi_0(0)}\propto\ket{\psi_0(0)},\\
  \label{eqa:cond2}
  &\ee^{-\cF\theta_T}\propto\I_2,
\end{align}
where $\theta_t=\int_0^t\omega(\tau)\dd{\tau}$,
$\ket{\psi_0(t)}=\ee^{-\ii\cH\theta_t}\ket{\psi_0(0)}$
is orthogonal to $\ket{\psi_1(t)}$ and $\ket{\psi_2(t)}$, and
$\cF=\qty[\cF_{ij}]_{i,j=1}^2$ with
\begin{equation}
  \cF_{ij}=-\ii\mel{\psi_i(t)}{\cH}{\psi_j(t)}
  =-\ii\mel{\psi_i(0)}{\ee^{\ii\cH\theta_t}
  \cH\ee^{-\ii\cH\theta_t}}{\psi_j(0)}
  =-\ii\mel{\psi_i(0)}{\cH}{\psi_j(0)}.
\end{equation}
Let
\begin{equation}
  \cH=\cE_0\ketbra{v_0}
  +\cE_1\ketbra{v_1}
  +\cE_2\ketbra{v_2}
  \label{eqa:H0}
\end{equation}
be the spectral decomposition of $\cH$ and
$\ket{\psi_0(0)}=a_0\ket{v_0}+a_1\ket{v_1}+a_2\ket{v_2}$. To ensure that the
quantum holonomy is nontrivial ($\Gamma(T)\not\propto\I_2$), at least two of $a_i$
should be nonzero. Without loss of generality, we assume that $a_1,a_2\ne 0$,
and furthermore $\cE_0=0$, then
\begin{equation}
  \ket{\psi_0(t)}=a_0\ket{v_0}
  +a_1\ee^{-\ii\cE_1\theta_t}\ket{v_1}
  +a_2\ee^{-\ii\cE_2\theta_t}\ket{v_2}.
  \label{eqa:psit}
\end{equation}
In the case that $a_0\ne 0$, Eq.~\eqref{eqa:cond1} implies that
$\ee^{-\ii\cE_1\theta_T}=\ee^{-\ii\cE_2\theta_T}=1$,
which would imply that $U(T)$ is trivial ($U(T)\propto\I_2$).
Therefore, we only consider the case that $a_0=0$, in which
we can choose $\ket{\psi_1(0)}=\ket{v_0}$ and
$\ket{\psi_2(0)}=a_2^*\ket{v_1}-a_1^*\ket{v_2}$.
Then, Eq.~\eqref{eqa:cond1} holds if and only if
$\ee^{\ii(\cE_2-\cE_1)\theta_T}=1$, i.e.,
\begin{equation}
  (\cE_2-\cE_1)\theta_T=2\pi N
  \label{eqa:cond1s}
\end{equation}
for some nonzero integer $N$ ($N=0$, i.e., $\theta_T=0$ or $\cE_1=\cE_2$
would result in a trivial $U(T)$ or $\Gamma(T)$).
In addition, we have $\cF_{11}=\cF_{12}=\cF_{21}=0$ and
$\cF_{22}=-\ii[\cE_1+(\cE_2-\cE_1)\abs{a_1}^2]$,
where we have used the normalization condition $\abs{a_1}^2+\abs{a_2}^2=1$.
Then Eq.~\eqref{eqa:cond2} holds if and only if
$[\cE_1+(\cE_2-\cE_1)\abs{a_1}^2]\theta_T=2\pi m$ for some integer $m$.
Together with Eq.~\eqref{eqa:cond1s}, we obtain that
\begin{equation}
  \abs{a_1}^2=\frac{m}{N}-\frac{\cE_1}{\cE_2-\cE_1}.
  \label{eqa:cond2s}
\end{equation}
Indeed, one can directly verify that the conditions
(\ref{eqa:cond1s},\,\ref{eqa:cond2s}) are sufficient for the two-dimensional
cyclic evolution $\qty{\ket{\psi_1(t)},\ket{\psi_2(t)}}$ ($0\le t\le T$)
being purely holonomic, under which
\begin{equation}
  U(T)=\Gamma(T)=\diag\qty(1,\,\ee^{-\ii\frac{2\pi N\cE_1}{\cE_2-\cE_1}}).
  \label{eqa:UT}
\end{equation}
Notably, in the widely-used nonadiabatic holonomic quantum computation, both
the original case \cite{Sjoeqvist.etal2012,Xu.etal2012} and single-shot case
\cite{Xu.etal2015,Sjoeqvist2016}, a special case ($m=0$) of
Eq.~\eqref{eqa:cond2s} is employed for implementing single-qubit quantum gates.

\end{document}